\pdfoutput=1

\documentclass[11pt,twoside,a4paper,cmspaper,final,collab]{cms-tdr}
\def\svnVersion{6fd9f3b-D}\def\svnDate{2020/09/21}\def\cmsCernNoTag{CERN-EP-2020-146}\def\cmsCernDate{\today}\def\cmsMessage{"Published in Physical Review D as \href{http://dx.doi.org/10.1103/PhysRevD.102.092007}{\doi{10.1103/PhysRevD.102.092007}.}"}

\begin{document}\cmsNoteHeader{BPH-19-001}

\hyphenation{had-ron-i-za-tion}
\hyphenation{cal-or-i-me-ter}
\hyphenation{de-vices}

\newlength\cmsFigWidth
\ifthenelse{\boolean{cms@external}}{\setlength\cmsFigWidth{0.98\columnwidth}}{\setlength\cmsFigWidth{0.60\textwidth}}
\newlength\cmsDoubleFigWidth
\ifthenelse{\boolean{cms@external}}{\setlength\cmsDoubleFigWidth{0.98\columnwidth}}{\setlength\cmsDoubleFigWidth{0.48\textwidth}}
\newlength\cmsTabSkip\setlength\cmsTabSkip{10pt}
\ifthenelse{\boolean{cms@external}}{\providecommand{\cmsLeft}{top\xspace}}{\providecommand{\cmsLeft}{left\xspace}}
\ifthenelse{\boolean{cms@external}}{\providecommand{\cmsRight}{bottom\xspace}}{\providecommand{\cmsRight}{right\xspace}}

\newcommand{\Bc}{\ensuremath{\BC^{\!\!\!+}}\xspace}
\newcommand{\BcStar}{\ensuremath{\BC^{\!\!\!*+}}\xspace}
\newcommand{\BcBoth}{\ensuremath{\BC^{\!\!(*)+}}\xspace}

\newcommand{\BcPrimeBoth}{\ensuremath{\BC\!\!^{(*)}\cmsSymbolFace{(2S)}^{\!+}}\xspace}
\newcommand{\BcPrime}{\ensuremath{\BC\cmsSymbolFace{(2S)}^{\!+}}\xspace}
\newcommand{\BcPrimeStar}{\ensuremath{\BC\!\!^{*}\cmsSymbolFace{(2S)}^{\!+}}\xspace}

\newcommand{\Bcpipi}{\ensuremath{\Bc \Pgpp \Pgpm}\xspace}
\newcommand{\BcStarpipi}{\ensuremath{\BcStar \Pgpp \Pgpm}\xspace}
\newcommand{\BcpipiBoth}{\ensuremath{\BcBoth \Pgpp \Pgpm}\xspace}

\newcommand{\BctoJpsipi}{\ensuremath{\Bc \! \to \! \cPJgy \, \Pgpp}\xspace}
\newcommand{\BctoJpsiK}{\ensuremath{\Bc \! \to \! \cPJgy \, \PK^{+}}\xspace}
\newcommand{\BctoJpsiX}{\ensuremath{\Bc \! \to \! \cPJgy \, \Pgpp \, X}\xspace}

\newcommand{\Jpsipi}{\ensuremath{\cPJgy \, \Pgpp}\xspace}
\newcommand{\psimumu}{\ensuremath{\cPJgy \! \to \! \MM}\xspace}

\newcommand{\BcPrimeBothtoBcpipi}{\ensuremath{\BcPrimeBoth \!\to\! \BC\!\!^{(*)+} \Pgpp \Pgpm}\xspace}

\title{\texorpdfstring
{Measurement of \BcPrime and \BcPrimeStar cross section ratios\\
in proton-proton collisions at $\sqrt{s} = 13$\TeV}
{Measurement of Bc+(2S) and Bc*+(2S) cross section ratios
in proton-proton collisions at sqrts = 13 TeV}}

\date{\today}

\abstract{
The ratios of the \BcPrime to \Bc, \BcPrimeStar to \Bc, and  \BcPrimeStar to \BcPrime production cross sections are measured in proton-proton collisions at $\sqrt{s} = 13$\TeV,  using a data sample collected by the CMS experiment at the LHC, corresponding to an integrated luminosity of 143\fbinv. The three measurements are made in the \Bc meson phase space region  defined by the transverse momentum $\pt > 15$\GeV and absolute rapidity $\abs{y} < 2.4$, with the excited \BcPrimeBoth states reconstructed  through the \BcpipiBoth, followed by the \BctoJpsipi and \psimumu decays. The \BcPrime to \Bc, \BcPrimeStar to \Bc, and \BcPrimeStar to \BcPrime cross section ratios,  including the unknown \BcPrimeBothtoBcpipi branching fractions, are  $(3.47 \pm 0.63\stat \pm 0.33\syst)\%$, $(4.69 \pm 0.71\stat \pm 0.56\syst)\%$, and  $1.35 \pm 0.32\stat \pm 0.09\syst$, respectively. None of these ratios shows a significant dependence on the \pt or $\abs{y}$ of the \Bc meson. The normalized dipion invariant mass distributions from the decays \BcPrimeBothtoBcpipi are also reported.
}

\hypersetup{%
pdfauthor={CMS Collaboration},%
pdftitle={Measurement of Bc+(2S) and Bc*+(2S) cross section ratios in proton-proton collisions at sqrts = 13 TeV},%
pdfsubject={CMS},%
pdfkeywords={CMS, physics, excited Bc mesons, production, properties}}

\maketitle

\section{Introduction}

The production cross sections of the \Bc family of mesons,
quark-antiquark bound states of two different flavors, charm and beauty, 
are significantly smaller than those of the charmonium and bottomonium states.
The unprecedented collision energies and integrated luminosities of the 
proton-proton ($\Pp\Pp$) data samples collected at the CERN LHC allow,
for the first time, detailed studies regarding the production and properties of \Bc quarkonia.
The observation of the \BcPrime and \BcPrimeStar states 
was recently reported by the CMS experiment~\cite{Bc2sObservationCMS},
using a $\Pp\Pp$ data sample collected at $\sqrt{s} = 13$\TeV between 2015 and 2018,
on the basis of well-resolved peaks in the \Bcpipi invariant mass distribution, 
with the \Bc meson reconstructed in the \BctoJpsipi decay channel, and \psimumu.
The LHCb Collaboration also reported the observation of the \BcPrimeStar state, 
using a $\Pp\Pp$ data sample collected at 7, 8, and 13\TeV~\cite{Bc2sObservationLHCb}.
Masses of the \BcPrime and \BcPrimeStar  states are found to be consistent
with theoretical predictions~\cite{Gregory:2009hq,Dowdall:2012ab,Mathur:2018epb}.
These results stimulated new theoretical studies aimed at reaching a better understanding of the 
\Bc quarkonium family, such as those reported in Refs.~\cite{Eichten:2019gig,Berezhnoy:2019yei}.

The present paper reports an analysis that complements the previous
observation of the \BcPrime and \BcPrimeStar states~\cite{Bc2sObservationCMS}
with the measurement of the \BcPrime to \Bc, \BcPrimeStar to \Bc, and \BcPrimeStar to \BcPrime cross section ratios,
an important step in making further progress on understanding these two excited \Bc states.
The invariant mass distributions of the pair of pions emitted in the $\BcPrimeBoth \to \BcpipiBoth$ decays are also presented,
to probe the existence of possible intermediate structure analogous to the ones observed in decays between the 2S and 1S states
of charmonium and bottomonium~\cite{Eichten:2019gig,Berezhnoy:2019yei}.
Throughout this paper, 
\BcBoth denotes \Bc or \BcStar, and \BcPrimeBoth denotes \BcPrime or \BcPrimeStar. 
Charge-conjugate states are also implied, unless stated otherwise.
The data sample of 13\TeV $\Pp\Pp$ collisions used in this analysis
corresponds to an integrated luminosity of 143\fbinv and was collected by CMS between 2015 and 2018.
The measurements are performed in a phase space region defined by the \Bc meson transverse momentum
$\pt > 15$\GeV and rapidity $\abs{y} < 2.4$.

\section{Experimental apparatus, data sample, and event selection}

The central feature of the CMS apparatus is a superconducting solenoid of 6\unit{m} internal diameter,
providing a magnetic field of 3.8\unit{T}.
Within the solenoid volume are a silicon pixel and strip tracker, a lead tungstate crystal electromagnetic calorimeter,
and a brass and scintillator hadron calorimeter, each composed of a barrel and two endcap sections.
Muons are measured in the pseudorapidity range $\abs{\eta} < 2.4$, 
with detection planes made using three technologies: drift tubes, cathode strip chambers, and resistive plate chambers. 
Matching muons to tracks measured in the silicon tracker results in a relative transverse momentum resolution, 
for muons with \pt up to 100\GeV, of 1\% in the barrel and 3\% in the endcaps~\cite{bib:softmuon2}.
The single-muon trigger efficiency exceeds 90\% over the full $\eta$ range, 
and the efficiency to reconstruct and identify muons is greater than 96\%. 
A more detailed description of the CMS detector, together with a definition of the coordinate system used and
relevant kinematic variables, can be found in Ref.~\cite{bib:CMSJINST}.

The event sample was collected with a two-level trigger system~\cite{bib:CMStrigger}.
At level 1, custom hardware processors select events with two muons.
The high-level trigger requires an opposite-sign muon pair of invariant mass in the range 2.9--3.3\GeV,
a dimuon vertex fit $\chi^2$ probability larger than 10\%,
a distance of closest approach between the two muons smaller than 0.5\unit{cm},
and a distance between the dimuon vertex and the beam axis, $L_{xy}$, larger than three times its uncertainty.
Both muons must have $\pt > 4$\GeV and $\abs{\eta} < 2.5$.
In addition $\vec{\pt}$ must be aligned with the dimuon transverse decay displacement vector $\vec{L}_{xy}$ by requiring 
$\cos\theta > 0.9$, where ${\cos\theta} = {\vec{L}_{xy}\cdot \vec{\pt} / (L_{xy} \, \pt)}$.
The trigger also requires a third track in the event, compatible with being produced at the dimuon vertex (normalized $\chi^2 < 10$),
and  having $\pt > 1.2$\GeV,  $\abs{\eta} < 2.5$, and a significance on the track impact parameter of at least 2.
The offline reconstruction requires two opposite-sign muons matching those that triggered the detector readout,
with some requirements being stricter than at the trigger level,
such as $\abs{\eta} < 2.4$ and $\cos\theta > 0.98$.
The muon candidates must pass high-purity track quality requirements~\cite{TRK-11-001}, and
fulfill the soft-muon identification requirements~\cite{bib:softmuon2},
which imply, in particular, that there are more than five hits in the silicon tracker, with at least one in the pixel layers.
The two muons must also be close to each other in angular space:
$\sqrt{\smash[b]{(\Delta\eta)^2 + (\Delta\phi)^2}} < 1.2$, where $\Delta\eta$ and $\Delta\phi$ are
the differences in pseudorapidity and azimuthal angle, respectively, between their momenta.

\section{Measurement of the cross section ratios}

\subsection{Introduction}

The ratios of the \BcPrimeBoth to \Bc and \BcPrimeStar to \BcPrime cross sections, $R^{*+}$, $R^{+}$, and $R^{*+}/R^{+}$, respectively,
reported in this paper
are derived from the ratios of the measured yields, corrected by the detection efficiencies, $\epsilon$:
\begin{linenomath}
\begin{equation}
\begin{aligned}
\ifthenelse{\boolean{cms@external}}{
R^{+} \equiv & \,
\frac{\sigma(\BcPrime)} {\sigma(\Bc)} \mathcal{B}(\BcPrime \to \Bcpipi) \\
= & \, \frac{N(\BcPrime)} {N(\Bc)} \, \frac{\epsilon(\Bc)} {\epsilon(\BcPrime)},\\
R^{*+} \equiv & \,
\frac{\sigma(\BcPrimeStar)} {\sigma(\Bc)} \mathcal{B}(\BcPrimeStar \to \BcStarpipi) \\
= & \, \frac{N(\BcPrimeStar)} {N(\Bc)} \, \frac{\epsilon(\Bc)} {\epsilon(\BcPrimeStar)},\\
R^{*+} / R^{+} = & \,
\frac{\sigma(\BcPrimeStar)} {\sigma(\BcPrime)}
\frac{\mathcal{B}(\BcPrimeStar \to \BcStarpipi)} {\mathcal{B}(\BcPrime \to \Bcpipi)} \\
= & \, \frac{N(\BcPrimeStar)} {N(\BcPrime)} \, \frac{\epsilon(\BcPrime)} {\epsilon(\BcPrimeStar)}.
}{
R^{+} \equiv & \,
\frac{\sigma(\BcPrime)} {\sigma(\Bc)} \mathcal{B}(\BcPrime \to \Bcpipi) = 
\frac{N(\BcPrime)} {N(\Bc)} \, \frac{\epsilon(\Bc)} {\epsilon(\BcPrime)},\\
R^{*+} \equiv & \,
\frac{\sigma(\BcPrimeStar)} {\sigma(\Bc)} \mathcal{B}(\BcPrimeStar \to \BcStarpipi) = 
\frac{N(\BcPrimeStar)} {N(\Bc)} \, \frac{\epsilon(\Bc)} {\epsilon(\BcPrimeStar)},\\
R^{*+} / R^{+} = & \,
\frac{\sigma(\BcPrimeStar)} {\sigma(\BcPrime)} 
\frac{\mathcal{B}(\BcPrimeStar \to \BcStarpipi)} {\mathcal{B}(\BcPrime \to \Bcpipi)} = 
\frac{N(\BcPrimeStar)} {N(\BcPrime)} \, \frac{\epsilon(\BcPrime)} {\epsilon(\BcPrimeStar)}.
}
\label{eq:ratios}
\end{aligned}
\end{equation}
\end{linenomath}
The $\mathcal{B}$ parameters are the unknown branching fractions of the 
$\BcPrimeBoth \to \BcpipiBoth$ decays.
The \BcStar meson is assumed to decay to the \Bc ground state and a low-energy photon with a branching fraction of
100\%, where the photon is not reconstructed.

\subsection{Measurement of the \texorpdfstring{\Bc}{Bc} yield}

The \BctoJpsipi candidates are reconstructed through a kinematic vertex fit, 
combining the dimuon with another track.
The dimuon invariant mass is constrained to the world-average \cPJgy\ mass~\cite{PDG2018}
and the other track, assumed to be a pion, must fulfil $\abs{\eta} < 2.4$ and $\pt > 3.5$\GeV.
The primary vertex (PV) associated with the \Bc candidate is selected among 
all the reconstructed vertices~\cite{Fruhwirth:2007hz}
as the one with the smallest angle between the reconstructed \Bc momentum 
and the vector joining the PV with the \Bc decay vertex.
To avoid biases, this PV is then refitted without the tracks associated with the muons and the pion.
The \Bc candidates are required to have $\pt >15$\GeV, $\abs{y} < 2.4$, 
a kinematic vertex fit $\chi^2$ probability larger than 10\%, and 
a decay length (distance between the \Jpsipi vertex and the PV) larger than 100\mum.
If several \Bc candidates are found in the same event, which happens in 1.6\% of the events, only the
one with the highest \pt is kept. Simulation studies show that this choice identifies the correct
candidate with 99\% probability.
These selection criteria were defined through studies of simulated signal samples
and measured sideband events~\cite{Bc2sObservationCMS}.

\begin{figure}[h]
\centering
\includegraphics[width=\cmsFigWidth]{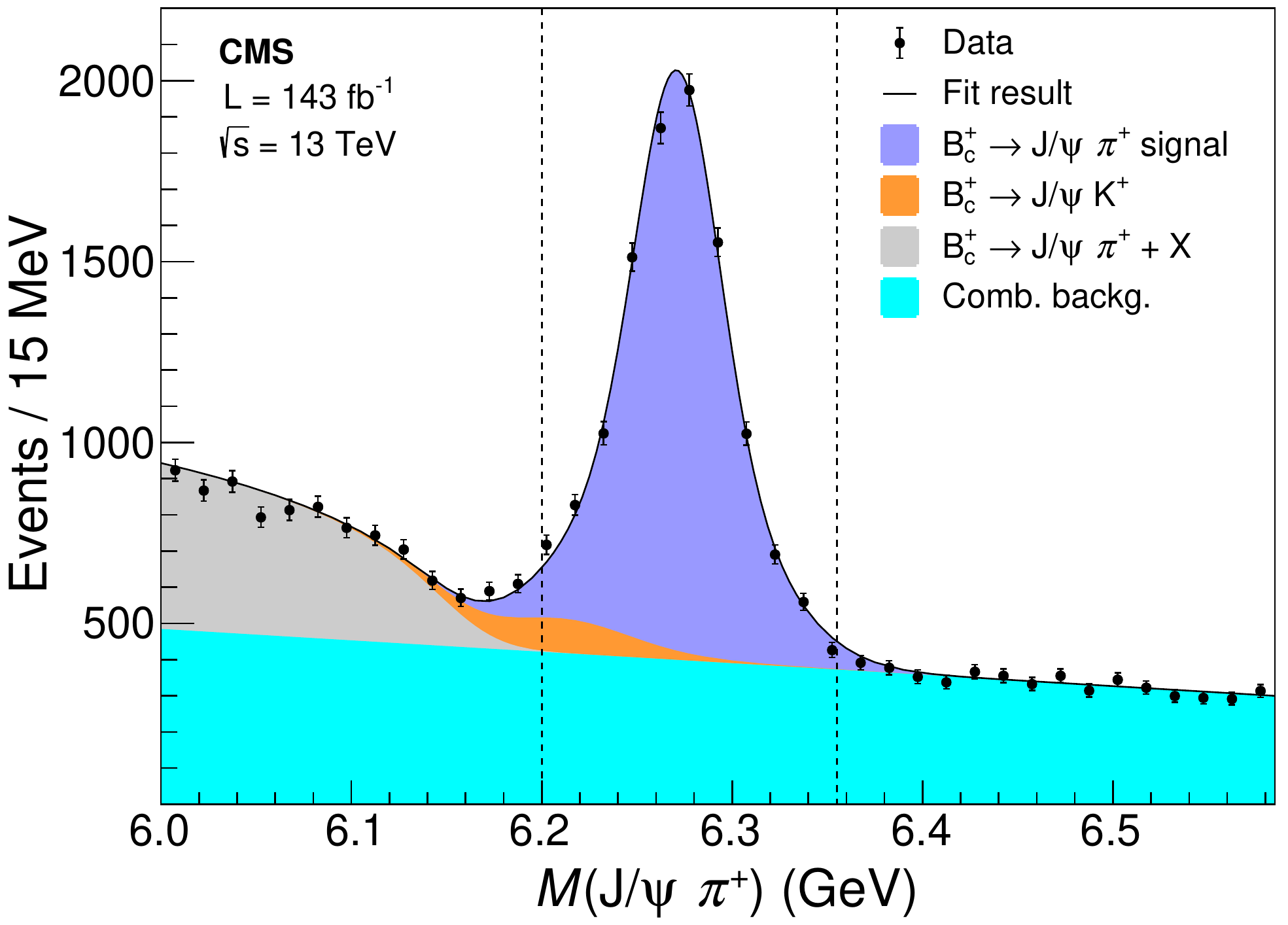}
\caption{Invariant mass distribution of the \BctoJpsipi candidates, after applying all event selection criteria~\cite{Bc2sObservationCMS}.
The fitted contributions are shown by the stacked distributions, the solid line representing their sum.
The vertical dashed lines indicate the mass window used to select the \Bc candidates for the
\BcPrimeBoth reconstruction.}
\label{fig:BcSignal}
\end{figure}

Figure~\ref{fig:BcSignal} shows the invariant mass distribution of the 
reconstructed and selected \BctoJpsipi candidates,
where the \Bc signal is clearly seen as a prominent peak~\cite{Bc2sObservationCMS}.
The result of an unbinned maximum-likelihood fit is also shown,
together with the signal and background contributions.
The underlying background is modeled as the sum of three terms:
(a) uncorrelated \JPsi-track combinations (combinatorial background), parametrized by a first-order polynomial;
(b) partially reconstructed \BctoJpsiX decays, only relevant for invariant mass values below 6.2\GeV
and parametrized by a generalized ARGUS function~\cite{Albrecht:1990am} 
convolved with a Gaussian resolution;
and (c) a small contribution from \BctoJpsiK decays, 
with a shape fixed from simulation studies (described later)
and a normalization fixed by the \BctoJpsipi yield,
scaled by the ratio of the corresponding branching fractions~\cite{lhcb_JpsiK} and reconstruction efficiencies.
The \Bc signal peak is modeled by a double-Gaussian function,
\begin{linenomath}
\begin{equation}
\label{eq:doubleG}
w G(\mu,\sigma_1) + (1-w) G(\mu,\sigma_2),
\end{equation}
\end{linenomath}
where $G(\mu, \sigma)$ represents a Gaussian function with mean $\mu$ and standard  deviation $\sigma$, and $w$ is 
the relative fraction of the narrower Gaussian in the fit. The single mean $\mu$ corresponds to the average reconstructed \Bc mass.
The fit gives $w = 47$\%, $\sigma_1 = 21$\MeV, and $\sigma_2 = 42$\MeV,
the very different Gaussian widths reflecting the fact that 
the \Bc mass resolution depends on rapidity,
degrading from the barrel to the endcap regions.
The \Bc mass resolution~\cite{Bc2sObservationCMS} agrees with expectations from simulation studies, of approximately 34\MeV.

The fitted \Bc mass is $M(\Bc) = 6271.1 \pm 0.5\MeV$ and
the \Bc signal yield is $7629 \pm 225$ events, 
where the uncertainties are statistical only.
The measured invariant mass distribution is well reproduced by the sum of the fitted contributions,
reflected in the  $\chi^2$ between the binned distribution and the fit function of 
35 for 30 degrees of freedom.

\subsection{Measurement of the \texorpdfstring{\BcPrime}{Bc+(2S)} and \texorpdfstring{\BcPrimeStar}{Bc*+(2S)} yields}

The \BcPrime and \BcPrimeStar candidates are also reconstructed through vertex kinematic fits,
combining a \Bc candidate with two opposite-sign, high-purity tracks, assumed to be pions.
The selected \Bc candidates must have invariant mass in the 6.2--6.355\GeV range,
where the low-mass edge is selected so as to avoid the background caused by
partially reconstructed decays (represented by the gray area below 6.2\GeV in Fig.~\ref{fig:BcSignal}).
The lifetimes of the \BcPrime and \BcPrimeStar are assumed to be negligible 
with respect to the measurement resolution,
so that the production and decay vertices essentially coincide.
Therefore, the daughter pions are among the tracks used in the refitted PV.
Furthermore, one of the pions must have $\pt > 0.8$\GeV and the other $\pt > 0.6$\GeV.
The \Bcpipi candidates must have $\abs{y} < 2.4$ and 
a vertex kinematic fit $\chi^2$ probability larger than 10\%.
As before, if several \Bcpipi candidates are found in the same event, only the one with the highest \pt is kept.

\begin{figure}[htb]
\centering
\includegraphics[width=\cmsFigWidth]{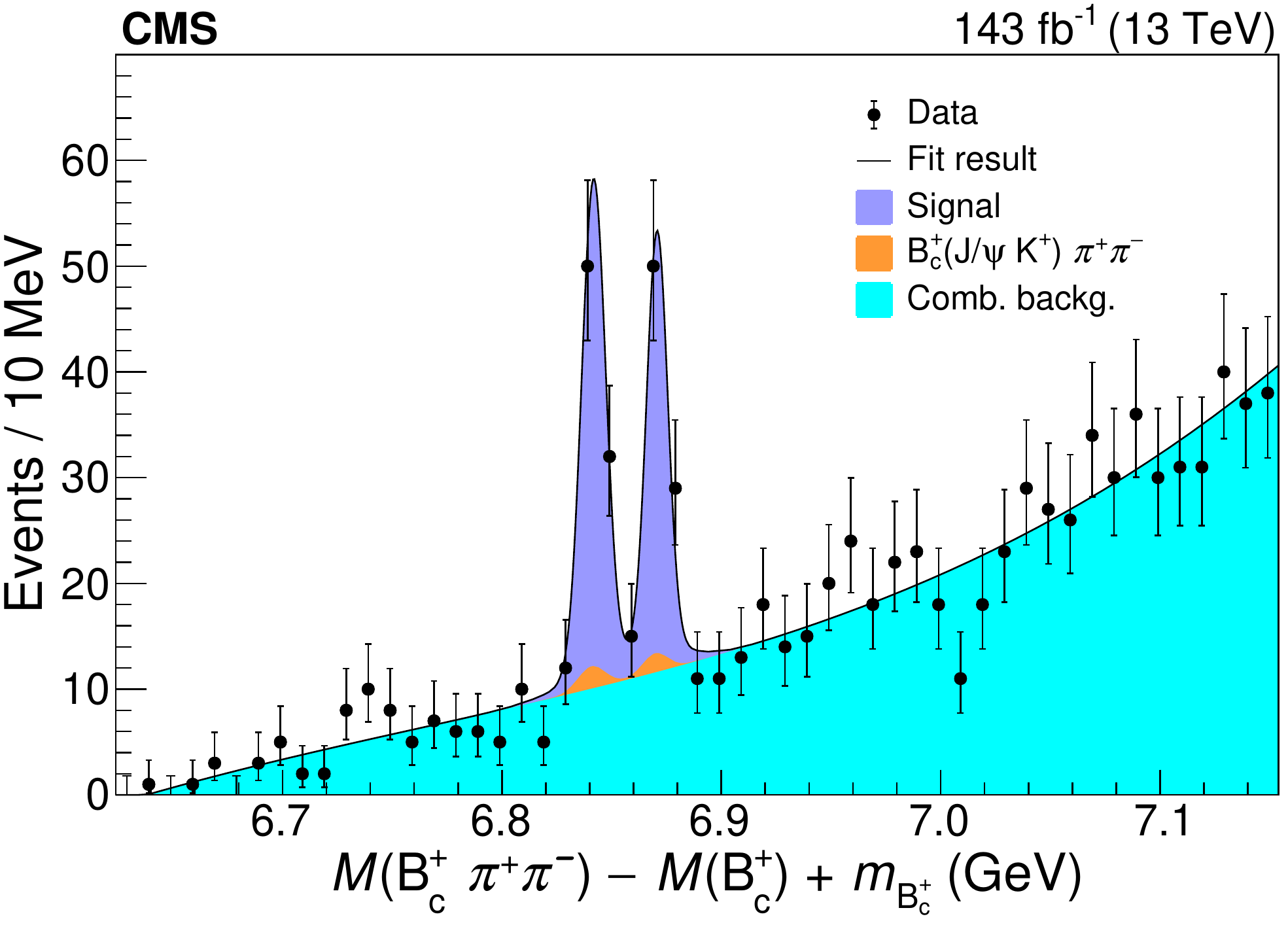}
\caption{Invariant mass distribution of the \BcPrimeBothtoBcpipi candidates~\cite{Bc2sObservationCMS}.
The \BcPrimeStar corresponds to the lower-mass peak, the \BcPrime to the higher.
The fitted contributions are shown by the stacked distributions, the solid line representing their sum.
}
\label{fig:Bc2sSignal}
\end{figure}

Figure~\ref{fig:Bc2sSignal} shows the $M(\Bcpipi) - M(\Bc) + m_{\Bc}$ distribution, 
where $M(\Bcpipi)$ and $M(\Bc)$ are the reconstructed invariant masses 
of the \Bcpipi and \Bc candidates, respectively, 
and $m_{\Bc}$ is the world-average \Bc mass~\cite{PDG2018}.
This variable is used in the analysis because it is measured with a better resolution than $M(\Bcpipi)$, 
given that some of the measurement uncertainties cancel in the difference.
The measured distribution is fitted to a superposition of two 
signal peaks using the same parametrization as in Eq.~\ref{eq:doubleG},
plus a third-order Chebyshev polynomial, modeling the nonpeaking, combinatorial background.
Two background contributions arising from \BctoJpsiK decays are also considered,
with shapes identical to those of the signal peaks,
ignoring a negligible shift (less than 1\MeV) to lower mass values,
and normalizations fixed by the ratio of the \BctoJpsiK to \BctoJpsipi signal yields.

Given the small number of events in the two signal peaks,
the $w$ and $\sigma_2$ double-Gaussian parameters are fixed to values determined in simulation studies:
$w = 92$\% and $\sigma_2 = 3.1 \, \sigma_1$ for the lower-mass peak; and
$w = 86$\% and $\sigma_2 = 2.8 \, \sigma_1$ for the higher-mass peak.
The two resonances are well resolved, 
with a mass difference of $28.9 \pm 1.5$\MeV,
where the uncertainty is statistical only.
The widths of the peaks are consistent with the measurement resolution 
evaluated through simulation studies, which is approximately $\sigma = 6$\MeV~\cite{Bc2sObservationCMS}.
The unbinned extended maximum-likelihood fit gives $67 \pm 10$ and $52 \pm 9$ events 
for the lower- and higher-mass peaks, respectively.
The quality of the fit can be quantified through the $\chi^2$ per degrees of freedom ratio, 41/35.

As explained in Ref.~\cite{Bc2sObservationCMS},
the \BcPrimeStar peak is seen in the \Bcpipi invariant mass distribution at a mass value 
lower than that of the \BcPrime peak.
The reason is that,
contrary to what happens to the \BcPrime, which decays directly to $\Bc \, \Pgpp \Pgpm$,
the \BcPrimeStar meson decays to \BcStarpipi where 
the photon emitted in the subsequent $\BcStar \! \to \! \Bc \, \gamma$ decay has too low energy to be reconstructed.
Therefore, the \BcPrimeStar peak is seen in the \Bcpipi mass spectrum
at the mass $M(\BcPrime) - \Delta M$, 
where $\Delta M \equiv [ M(\BcStar) - M(\Bc) ] - [ M(\BcPrimeStar) - M(\BcPrime)]$.
Since $M(\BcStar) - M(\Bc)$ is expected to be larger than $M(\BcPrimeStar) - M(\BcPrime)$,
the \BcPrimeStar state corresponds to the lower-mass peak~\cite{Gregory:2009hq,Dowdall:2012ab,Mathur:2018epb}.

\subsection{Reconstruction efficiencies}

With respect to the observation analysis reported in Ref.~\cite{Bc2sObservationCMS},
the main challenge in the determination of the \BcPrimeBoth to \Bc cross section ratios
is the evaluation of the corresponding (relative) detection efficiencies.
Since the trigger requires \psimumu from the \BctoJpsipi decay, the trigger efficiencies for the \Bc and 
\Bcpipi candidates are essentially the same and cancel in the cross section ratios.
So only the reconstruction efficiencies need to be evaluated,
which is done using simulated event samples.
All three mesons (\Bc, \BcPrime, and \BcPrimeStar) are generated using the \textsc{bcvegpy}~2.2~\cite{Bcveg}
Monte Carlo event generator. 
The events are then passed to \PYTHIA~8.230~\cite{bib:Pythia} to simulate the hadronization process.
The decays are performed by the \EVTGEN~1.6.0 package~\cite{bib:EvtGen} and the 
quantum electrodynamic final-state radiation is modeled with \PHOTOS~3.61~\cite{bib:PHOTOS2}.
The simulated events are then processed through a detailed simulation of the CMS detector, 
based on the \GEANTfour package~\cite{geant4},
using the same trigger and reconstruction algorithms used to 
collect and process the data.
The simulated events include multiple $\Pp\Pp$ interactions in the same 
or nearby beam crossings (pileup), with a distribution matching the one observed in data.
Monte Carlo samples were extensive validated using control regions in data.

The \BcPrime and \BcPrimeStar efficiencies are computed as
$N_{\text{rec}}(\BcPrimeBoth) / N_{\text{gen}}(\BcPrimeBoth)$,
where $N_{\text{gen}}(\BcPrimeBoth)$ are the numbers of \BcPrimeBoth
events generated in the \BcpipiBoth channel,
in the phase space region of the analysis, $\pt(\Bc) > 15$\GeV and $\abs{y(\Bc)}<2.4$,
and $N_{\text{rec}}(\BcPrimeBoth)$ are the numbers of events that survive all the reconstruction steps
and event selection criteria.
The \Bc efficiency is computed in a completely analogous way, 
except that it uses \Bc events generated in the \BctoJpsipi decay channel.
These evaluations are independently made for 
the 2016, 2017, and 2018 running periods.
The events collected in 2015, corresponding to 2\% of the total sample,
are treated the same as the 2016 sample for the purpose of efficiency determination.
It was checked that the 2016 Monte Carlo simulation describes the 2015 data well enough so that no residual systematic uncertainty is required.
The final efficiencies are obtained as weighted averages,
using the integrated luminosities as weights:
$2.8+36.1$, 42.1, and 61.6\fbinv, respectively, for the $2015+2016$, 2017, and 2018 
periods~\cite{bib:CMS-PAS-LUM-15-001,bib:CMS-PAS-LUM-17-001,bib:CMS-PAS-LUM-17-004,bib:CMS-PAS-LUM-2018}.
The results are 
$\epsilon(\Bc) = 1.31$\%, 
$\epsilon(\BcPrime) = 0.26$\%, 
and $\epsilon(\BcPrimeStar) = 0.24$\%.
The \BcPrime and \BcPrimeStar reconstruction efficiencies are very similar,
the slightly smaller \BcPrimeStar value reflecting the (missed) low-energy photon, 
which implies a small reduction of the \Bcpipi phase space.

Table~\ref{tab:effs} lists the efficiency ratios relevant for the determination of the cross section ratios.
The first uncertainty (``Stat.") shown reflects the finite size of the three simulated samples.
The second (``Spread") 
reflects the standard deviation of the computed values around their average
and is used to conservatively cover potential residual mismatches 
between the running conditions and the settings used in simulation.
For example, it could be that the simulated samples do not accurately reproduce 
the time evolution of the instantaneous luminosity within each data-taking period,
which would create differences in the measured and simulated pileup distributions.
The last column (``Pions") reflects the uncertainty in the reconstruction efficiency~\cite{tracking}
of the two pions emitted in the $\BcPrimeBoth \to \BcpipiBoth$ decays. This uncertainty is relevant 
for the $R^{*+}$ and $R^{+}$ ratios, but cancels in the $R^{*+}/R^{+}$ ratio.

\begin{table}[ht]
\centering
\topcaption{Ratios of the reconstruction efficiencies relevant for the determination 
of the $R^{+}$, $R^{*+}$, and $R^{*+}/R^{+}$ cross section ratios.
The central values are followed by the several uncertainties presented in the text.}
\label{tab:effs}
\renewcommand{\arraystretch}{1.2}
\begin{scotch}{l c c c c}
           & Central & Stat.\ & Spread\ & Pions \\
\hline
$\epsilon(\BcPrime) / \epsilon(\Bc)$            & 0.196 & 1.1\% & 1.8\% & 4.2\% \\
$\epsilon(\BcPrimeStar) / \epsilon(\Bc)$        & 0.187 & 1.0\% & 1.6\% & 4.2\% \\
$\epsilon(\BcPrimeStar) / \epsilon(\BcPrime)$   & 0.955 & 1.4\% & 0.9\% & \NA   \\
\end{scotch}
\end{table}

\subsection{Determination of the cross section ratios}

Correcting the yield ratios by the corresponding efficiency ratios leads to the following 
\BcPrime to \Bc, \BcPrimeStar to \Bc, and \BcPrimeStar to \BcPrime cross section ratios,
always including the \BcPrimeBothtoBcpipi branching fractions, 
and always for $\pt(\Bc) > 15$\GeV and $\abs{y(\Bc)}<2.4$:
\begin{linenomath}
\begin{equation}
\begin{aligned}
R^{+} = & \, (3.47 \pm 0.63)\%,\\
R^{*+} = & \, (4.69 \pm 0.71)\%,\quad\text{and}\\
R^{*+} / R^{+} = & \, 1.35 \pm 0.32.
\label{eq:results}
\end{aligned}
\end{equation}
\end{linenomath}
The quoted uncertainties are statistical only. 
The fact that the \BcPrimeBoth events are a subset of the \Bc events has a negligible effect 
(less than 1\%) on the uncertainties.
The correlation between \BcPrimeStar and \BcPrime yields, used in the double cross section ratio, 
is taken into account using an alternative fit to the $M(\Bcpipi) - M(\Bc) + m_{\Bc}$ distribution, 
which directly provides the ratio of these yields.
It is worth noting again that these ratios include branching fractions (shown in Eq.~(\ref{eq:ratios})) 
that have not yet been measured.

\subsection{Dependence on the \texorpdfstring{\Bc}{Bc} kinematics}

In order to probe if these cross section ratios show a dependence on the kinematics of the
\Bc meson, the analysis is redone after splitting the events into three \Bc meson \pt bins and 
(independently) into three $\abs{y}$ bins. The bin edges are chosen so as to have similar
uncertainties in the three bins: 15, 22.5, 30, and 60\GeV for \pt, and 0, 0.4, 0.8, and 2.4 for $\abs{y}$.
The amount of events with $\pt>60$\GeV corresponds to 3.4\% of the total sample and they are excluded
from these kinematical distributions.

\begin{figure}[htb]
\centering
\includegraphics[width=\cmsDoubleFigWidth]{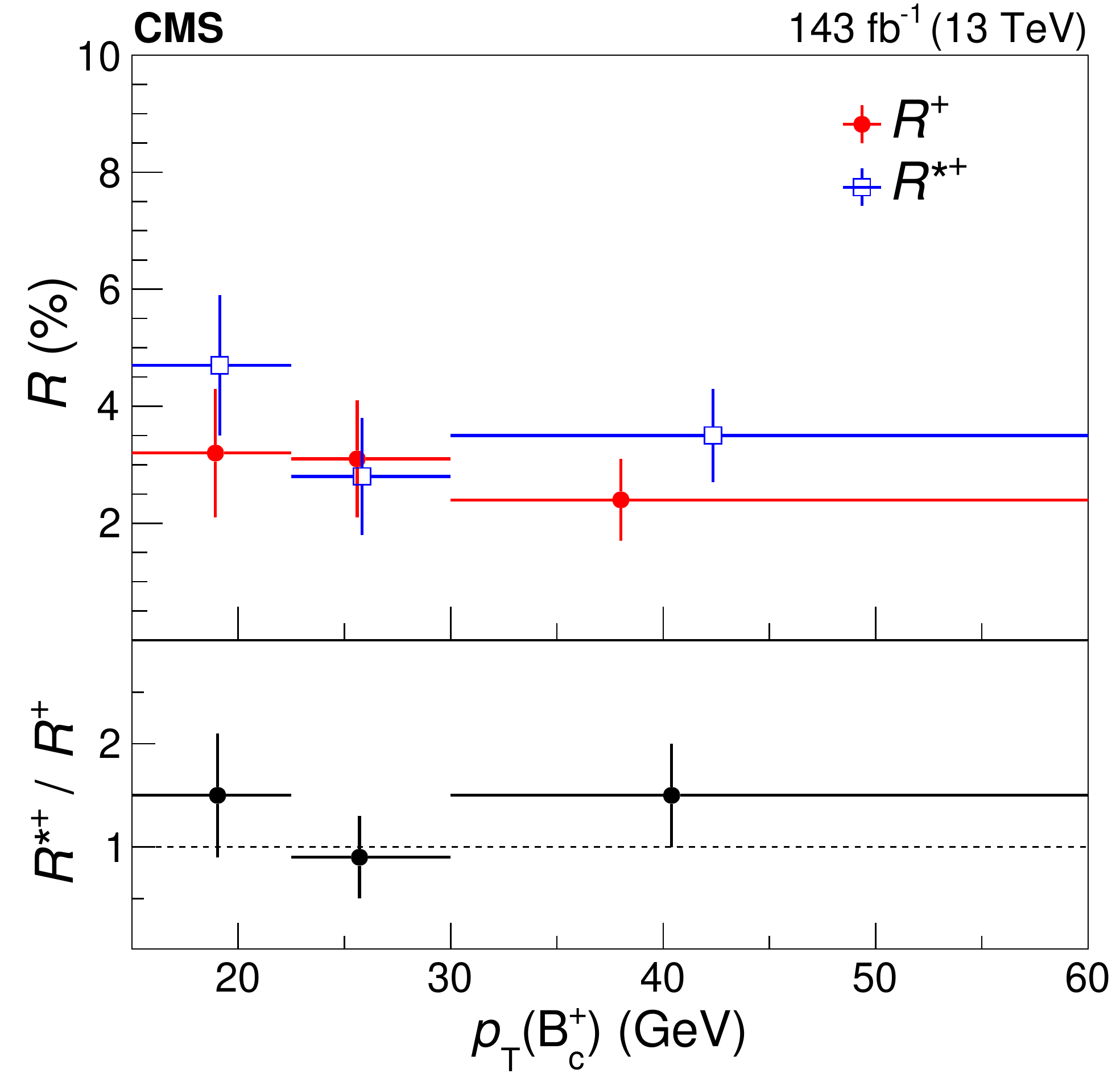}
\includegraphics[width=\cmsDoubleFigWidth]{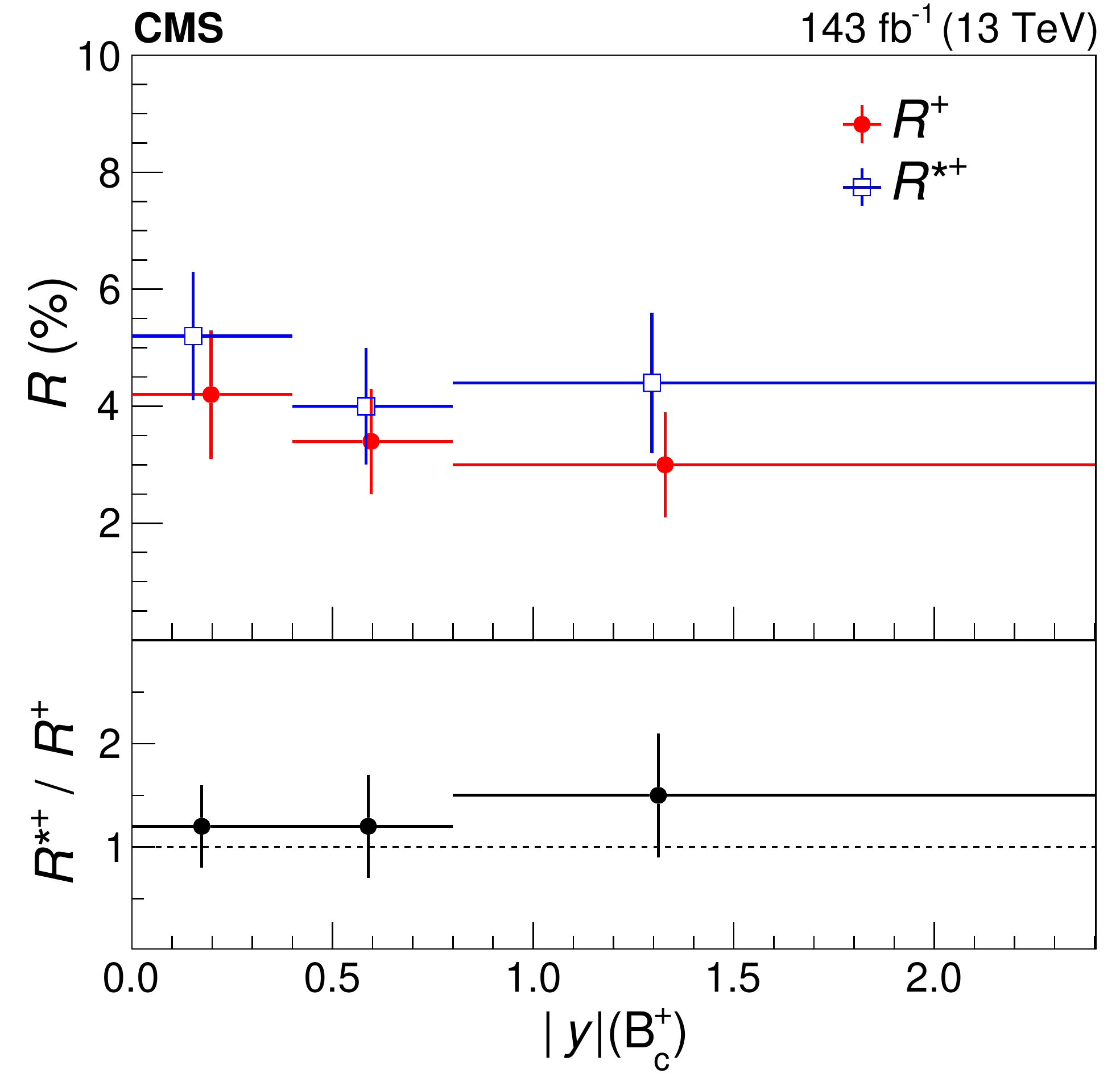}
\caption{The $R^{+}$ and $R^{*+}$ (upper), and $R^{*+}/R^{+}$ (lower) 
cross section ratios, including the \BcPrimeBothtoBcpipi branching fractions,
as functions of the \Bc \pt (\cmsLeft) and $\abs{y}$ (\cmsRight).
The horizontal bars show the bin widths.
The markers are shown at the average \Bc \pt or $\abs{y}$ values of the 
events contributing to each bin, in the background-subtracted distributions,  
and the vertical bars represent the statistical uncertainties only. The
systematic uncertainties are essentially independent of the \Bc kinematics.}
\label{fig:pT_y}
\end{figure}

As shown in Fig.~\ref{fig:pT_y}, none of the measured ratios shows significant variations
with the \pt or $\abs{y}$ of the \Bc meson, within the probed kinematical regions.
The markers are shown at the average \Bc \pt or $\abs{y}$ values of the events contributing to each bin.
The horizontal displacements between the markers seen in the top panels reflect the differences between the \BcPrime and \BcPrimeStar
kinematic distributions.

Reporting the cross section ratios as a function of the \Bc kinematics and in a phase space domain defined by the \Bc is the
choice that best reflects the data analysis procedure and that cancels to the largest extent the systematic uncertainties
related to the \Bc detection. Given the relatively small mass difference between the mother \BcPrimeBoth  and the daughter
\Bc states, the ratio of laboratory momentum to mass remains practically unchanged in the decays, on average, so that
the following kinematical relations hold to a very good approximation: $y^M = y^d$ and $\pt^M = (M/m) \, \pt^d$, where
$y^M$, $\pt^M$, and $M$ (respectively $y^d$, $\pt^d$, and $m$) are the rapidity, \pt, and mass of the mother (respectively daughter)~\cite{Faccioli:2017hym}.

\subsection{Systematic uncertainties}

Several sources of systematic effects 
that could potentially affect the measurement of the cross section ratios have been considered.
For each of those effects, the analysis has been redone using an alternative option
and the resulting cross section ratios are compared to those obtained in the baseline analysis.
The observed difference between the two results 
is taken as the systematic uncertainty associated with that specific effect.

Naturally, no uncertainties are considered in factors that affect identically the numerator and denominator values 
that provide the cross section ratios, such as the efficiency of the \JPsi trigger used to collect the event sample
or the efficiency of the event selections that determine the total number of \BctoJpsipi candidates
contributing to Fig.~\ref{fig:BcSignal}.
But even if the integral of the measured \Jpsipi invariant mass distribution does not change, 
it is possible to vary the extracted \Bc yield by changing the functions used in the fit
to describe the shapes of the signal and background contributions,
given that such variations might change the assignment of some events 
from the \Bc yield to the background yield, or vice versa.
The importance of this effect is evaluated by independently varying
the signal and background models used in the fit.

The background model is varied by using an exponential function, 
instead of a first-order polynomial, to describe the uncorrelated \Jpsipi pairs.
The varied scenario for the \Bc signal line shape consisted in replacing the double-Gaussian function
by a Student's $t$ function~\cite{Studentt}.
Since these two variations only change the fitted \Bc yield, 
having no effect on the number of \BctoJpsipi candidates 
used in the search for the \BcPrimeBoth excited states, 
the corresponding (relative) systematic uncertainties, 
4.3\% for the signal model and 3.5\% for the background model, 
are identical for the $R^{+}$ and $R^{*+}$ ratios,
and cancel in the $R^{*+}/R^{+}$ double ratio.

The measurement of the \BcPrime and \BcPrimeStar yields is also affected by the choices
made to model the shapes of the signal peaks and the underlying combinatorial background
seen in Fig.~\ref{fig:Bc2sSignal}.
The effect of the signal modeling is evaluated with two independent approaches.
First, the default double-Gaussian function,
having a common mean and fixing the relative widths and amplitudes from fits to the simulated distributions,
is replaced by a single-Gaussian function.
The number of free parameters for each signal peak remains at three, 
but this simpler model is unable to describe the non-Gaussian tails of the peaks.
Second, the signal yields are evaluated with a simple procedure that avoids fitting the 
mass region of the two signal peaks, thereby being insensitive to specific signal shape models. 
It starts by fitting the signal-free mass sidebands with the background function and then
integrating that function within the two signal regions to evaluate the background yields under the peaks,
which are then subtracted from the total number of events in those two regions.
To evaluate the impact of the background model, these alternative fits have been made with
the third-order Chebyshev polynomial used in the baseline analysis and also 
with the function $\delta^\lambda \exp(\nu \, \delta)$, where 
$\delta \equiv M(\Bcpipi) - q_0$, and $\lambda$, $\nu$, and $q_0$ are free parameters.
Comparing the cross section ratios obtained using the alternative fits 
with those of the baseline fit leads to fit modeling systematic uncertainties of
5.9, 2.9, and 2.9\%,
respectively for the $R^{+}$, $R^{*+}$, and $R^{*+}/R^{+}$ ratios.

The fit of the \Bcpipi invariant mass distribution also includes two small contributions representing 
the cases where the \Bc meson decays through the \BctoJpsiK channel rather than through the
\BctoJpsipi channel assumed in the reconstruction. 
In the baseline analysis, 
these terms are modeled using the same shapes as the \BcPrimeBoth signal shapes 
and yields fixed to the yields of those resonances, 
scaled by the ratio of the two branching fractions, $0.079 \pm 0.008$~\cite{lhcb_JpsiK},
and by the ratio of the two reconstruction efficiencies, $1.06 \pm 0.01$, in the signal region defined above.
To evaluate the influence of these terms on the measured cross section ratios,
the analysis is redone varying those two scale factors by their uncertainties.
The results are insensitive to those variations, 
so no systematic uncertainty is assigned to this source.

When searching for \BcPrimeBoth candidates, the baseline analysis starts from an event sample
composed of \BctoJpsipi events with invariant mass in the 6.2--6.355\GeV range. 
In order to probe if a potential residual contribution of the partially reconstructed \Bc decays could have 
a significant effect on the determination of the cross section ratios, 
the analysis is repeated with the lowest allowed invariant mass value changed
from 6.2 to 6.1\GeV.
The results remain essentially identical, the variations being smaller than their statistical uncertainties, 
evaluated taking into account that one event sample is a subset of the other, 
so that the results are fully correlated.
Therefore, no systematic uncertainty is assigned to this potential effect.

The uncertainties affecting the ratios of reconstruction efficiencies already
presented in Table~\ref{tab:effs} translate directly into corresponding 
systematic uncertainties in the cross section ratios.
In the evaluation of the \BcPrimeBoth reconstruction efficiencies,
it is assumed that the two pions emitted in the \Bcpipi decay 
have no kinematical correlations between them,
besides the constraint of being decay products of the same mother particle.
To evaluate the sensitivity of the measured cross section ratios to this assumption,
the reconstruction efficiencies are recomputed under two other scenarios.
These assume that the $\Pgpp\Pgpm$ kinematic distributions (a) reflect the existence of an intermediate resonance, 
or (b) are dependent on the (different) spins  of the \BcPrime and \BcPrimeStar states.  The first scenario is simulated 
by independently reweighting the generated \BcPrimeBoth event samples, which previously reflected a simple phase space model, 
so that their $\Pgpp\Pgpm$ invariant mass distributions (``decay kinematics") match that in the data (presented in Section~\ref{sec:four}). 
The second scenario follows an analogous procedure using the helicity angle distribution (``helicity angle"), 
where the helicity angle is the angle between the directions of the \Pgpp and \Bc in the dipion rest frame.
The differences between the resulting ratios of reconstruction efficiencies and those
obtained in the baseline scenario are considered as systematic uncertainties:
1.5, 6.9, and 4.2\% for the decay kinematics, and 
1.0, 6.0, and 3.5\% for the helicity angle, 
respectively, for the $R^{+}$, $R^{*+}$, and $R^{*+}/R^{+}$ ratios.

Several studies have been performed to verify the stability of the results with respect to the selection criteria,
including the threshold values used to select the daughter particles. The variations in the reported ratios were smaller
than the respective uncertainties, computed accounting for the correlation induced by the overlap of the baseline
and varied event samples, so that no corresponding systematic uncertainty has been considered.

All the values mentioned above are listed in Table~\ref{table:Systematics}, 
which also shows the total systematic uncertainties,
computed as the sum in quadrature of the individual terms.

\begin{table}[ht]
\centering
\topcaption{Relative systematic uncertainties (in \%) in the cross section ratios, 
including the \BcPrimeBothtoBcpipi branching fractions,
corresponding to the sources described in the text.
The total uncertainty is the sum in quadrature of the individual terms.}
\label{table:Systematics}
\renewcommand{\arraystretch}{1.1}
\begin{scotch}{l ccc}
& $R^{+}$ & $R^{*+}$ & $R^{*+}/R^{+}$ \\ \hline
\Jpsipi fit model            & 5.5 & 5.5 & \NA \\
\Bcpipi fit model            & 5.9 & 2.9 & 2.9 \\
Efficiencies: statistical uncertainty   & 1.1 & 1.0 & 1.4 \\
Efficiencies: spread among years        & 1.8 & 1.6 & 0.9 \\
Efficiencies: pion tracking             & 4.2 & 4.2 & \NA \\
Decay kinematics                        & 1.5 & 6.9 & 4.2 \\
Helicity angle                          & 1.0 & 6.0 & 3.5 \\
[\cmsTabSkip] 
Total                                   & 9.5 & 12.0 & 6.4 \\
\end{scotch}
\end{table}

\section{Invariant mass distribution of the dipion system}\label{sec:four}

As a complement to the measurement of the cross section ratios,
it is also interesting to measure the invariant mass distributions of the dipions 
emitted in the \Bcpipi decays of the two \BcPrimeBoth states.
In particular, comparing these distributions to those seen in the analogous 
$\Pgy\to\cPJgy\,\Pgpp\Pgpm$ and $\PgUb\to\PgUa\,\Pgpp\Pgpm$ decays
should provide relevant information to characterize the excited \Bc states
and their production processes~\cite{Eichten:2019gig,Berezhnoy:2019yei}.

\begin{figure}[htb]
\centering
\includegraphics[width=\cmsFigWidth]{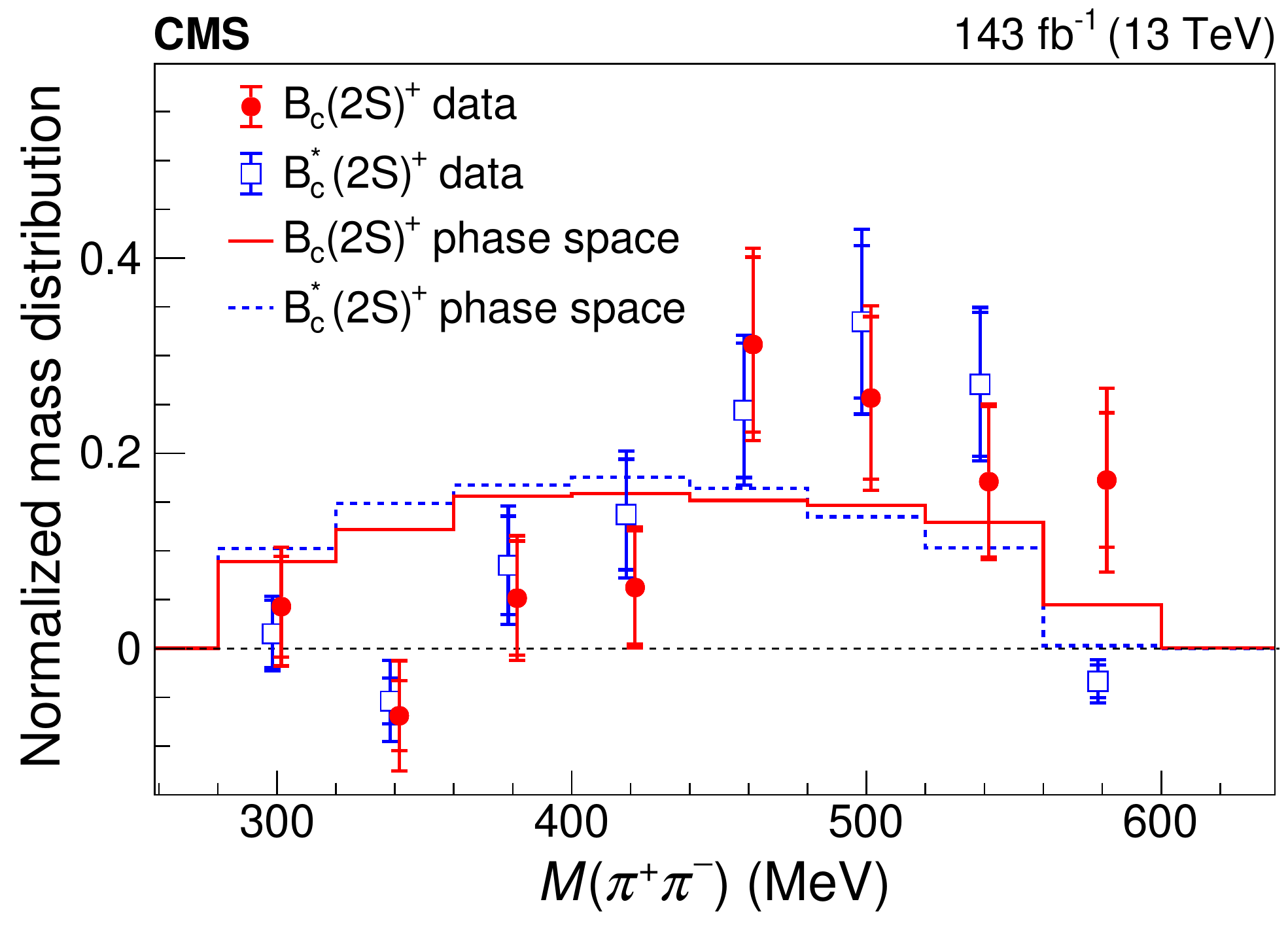}
\caption{
The dipion invariant mass distributions from \BcPrimeBothtoBcpipi decays in data, normalized to unity.
The inner and outer tick marks designate the statistical and total uncertainties, respectively. 
The lines show the corresponding predictions from phase space simulations.
}
\label{fig:mpipi_all}
\end{figure}

Figure~\ref{fig:mpipi_all} compares the invariant mass distributions, normalized to unity,
of the dipions emitted in the \BcPrime (closed red circles) and \BcPrimeStar (open blue squares) decays 
between themselves and with the two corresponding simulated phase space distributions (lines).
The \BcPrimeBoth data distributions are derived
from the \Bcpipi invariant mass distribution
shown in Fig.~\ref{fig:Bc2sSignal}.
The contribution of the background events under the peaks is subtracted 
using the shape of the measured same-sign dipion invariant mass spectrum and 
normalizing the sum of the $\Bc\Pgpp\Pgpp$ and $\Bc\Pgpm\Pgpm$ events
to the \Bcpipi spectrum in the invariant mass sideband regions.
The dipion invariant mass distributions have also been obtained using the sPlot technique~\cite{sPlot} to subtract the background, 
which resulted in distributions consistent with those reported in Fig.~\ref{fig:mpipi_all}.

Simulation studies show no dependence of the reconstruction efficiencies 
on the $\Pgpp\Pgpm$ invariant mass, so no correction is applied to these
normalized distributions, where only the shapes are informative.
For the same reason, systematic uncertainties that affect the distributions globally are
not relevant, as they have no impact on the shapes and are canceled by the normalizations.

The dipion mass-dependent systematic uncertainties have been evaluated by comparing,
bin by bin, the baseline distributions with those obtained in alternative analyses, where 
variations are made, as mentioned above, on the models used to fit the signal and background 
components of the \Bcpipi mass distribution and on the small contributions from the \BctoJpsiK 
and partially reconstructed \Bc decays.

As seen in Fig.~\ref{fig:mpipi_all}, the \BcPrimeBoth dipion invariant mass distributions are 
compatible with each other within the uncertainties, and have shapes different from the rather 
flat distributions predicted from the phase space simulations.

\section{Summary}

The ratios of the \BcPrime to \Bc, \BcPrimeStar to \Bc, and \BcPrimeStar to \BcPrime production cross sections,
$R^{+}$, $R^{*+}$, and $R^{*+}/R^{+}$, respectively, have
been measured in proton-proton collisions at $\sqrt{s} = 13$\TeV.
Data set used in the analysis corresponds to an integrated luminosity of 143\fbinv collected by the 
CMS experiment at the LHC between 2015 and 2018.

The \BcPrimeBoth mesons were reconstructed through the decays \BcPrimeBothtoBcpipi, followed by the \BctoJpsipi and \psimumu.
The measured cross section ratios, including the (unknown) \BcPrimeBothtoBcpipi branching fractions, are 
\begin{linenomath}
\begin{equation}
\begin{aligned}
R^{+} = & \, (3.47 \pm 0.63\stat \pm 0.33\syst)\%,\\
R^{*+} = & \, (4.69 \pm 0.71\stat \pm 0.56\syst)\%,\quad \text{and}\\
R^{*+} / R^{+} = & \, 1.35 \pm 0.32\stat \pm 0.09\syst.
\end{aligned}
\end{equation}
\end{linenomath}
No significant dependences on the transverse momentum \pt or rapidity $\abs{y}$ of the \Bc mesons have been observed 
for any of these three ratios.
The normalized dipion invariant mass distributions for the \BcPrimeBothtoBcpipi decays are also reported.
These results, obtained in the phase space region defined by \Bc meson $\pt > 15$\GeV 
and $\abs{y} < 2.4$, may provide new important input 
to improve the theoretical understanding of the nature of the \bbarc heavy-quarkonium states
and their production processes.

\begin{acknowledgments}
    We congratulate our colleagues in the CERN accelerator departments for the excellent performance of the LHC and thank the technical and administrative staffs at CERN and at other CMS institutes for their contributions to the success of the CMS effort. In addition, we gratefully acknowledge the computing centers and personnel of the Worldwide LHC Computing Grid for delivering so effectively the computing infrastructure essential to our analyses. Finally, we acknowledge the enduring support for the construction and operation of the LHC and the CMS detector provided by the following funding agencies: BMBWF and FWF (Austria); FNRS and FWO (Belgium); CNPq, CAPES, FAPERJ, FAPERGS, and FAPESP (Brazil); MES (Bulgaria); CERN; CAS, MoST, and NSFC (China); COLCIENCIAS (Colombia); MSES and CSF (Croatia); RIF (Cyprus); SENESCYT (Ecuador); MoER, ERC IUT, PUT and ERDF (Estonia); Academy of Finland, MEC, and HIP (Finland); CEA and CNRS/IN2P3 (France); BMBF, DFG, and HGF (Germany); GSRT (Greece); NKFIA (Hungary); DAE and DST (India); IPM (Iran); SFI (Ireland); INFN (Italy); MSIP and NRF (Republic of Korea); MES (Latvia); LAS (Lithuania); MOE and UM (Malaysia); BUAP, CINVESTAV, CONACYT, LNS, SEP, and UASLP-FAI (Mexico); MOS (Montenegro); MBIE (New Zealand); PAEC (Pakistan); MSHE and NSC (Poland); FCT (Portugal); JINR (Dubna); MON, RosAtom, RAS, RFBR, and NRC KI (Russia); MESTD (Serbia); SEIDI, CPAN, PCTI, and FEDER (Spain); MOSTR (Sri Lanka); Swiss Funding Agencies (Switzerland); MST (Taipei); ThEPCenter, IPST, STAR, and NSTDA (Thailand); TUBITAK and TAEK (Turkey); NASU (Ukraine); STFC (United Kingdom); DOE and NSF (USA).

    \hyphenation{Rachada-pisek} Individuals have received support from the Marie-Curie program and the European Research Council and Horizon 2020 Grant, contract Nos.\ 675440, 752730, and 765710 (European Union); the Leventis Foundation; the A.P.\ Sloan Foundation; the Alexander von Humboldt Foundation; the Belgian Federal Science Policy Office; the Fonds pour la Formation \`a la Recherche dans l'Industrie et dans l'Agriculture (FRIA-Belgium); the Agentschap voor Innovatie door Wetenschap en Technologie (IWT-Belgium); the F.R.S.-FNRS and FWO (Belgium) under the ``Excellence of Science -- EOS" -- be.h project n.\ 30820817; the Beijing Municipal Science \& Technology Commission, No. Z191100007219010; the Ministry of Education, Youth and Sports (MEYS) of the Czech Republic; the Deutsche Forschungsgemeinschaft (DFG) under Germany's Excellence Strategy -- EXC 2121 ``Quantum Universe" -- 390833306; the Lend\"ulet (``Momentum") Program and the J\'anos Bolyai Research Scholarship of the Hungarian Academy of Sciences, the New National Excellence Program \'UNKP, the NKFIA research grants 123842, 123959, 124845, 124850, 125105, 128713, 128786, and 129058 (Hungary); the Council of Science and Industrial Research, India; 
the Bilateral Scientific and Technological Cooperation Program between Italy and Mexico 2018-2020 (project MX18MO11 and additional MAECI project PGR 00783/2019);
the HOMING PLUS program of the Foundation for Polish Science, cofinanced from European Union, Regional Development Fund, the Mobility Plus program of the Ministry of Science and Higher Education, the National Science Center (Poland), contracts Harmonia 2014/14/M/ST2/00428, Opus 2014/13/B/ST2/02543, 2014/15/B/ST2/03998, and 2015/19/B/ST2/02861, Sonata-bis 2012/07/E/ST2/01406; the National Priorities Research Program by Qatar National Research Fund; the Ministry of Science and Higher Education, project no. 02.a03.21.0005 (Russia); the Programa Estatal de Fomento de la Investigaci{\'o}n Cient{\'i}fica y T{\'e}cnica de Excelencia Mar\'{\i}a de Maeztu, grant MDM-2015-0509 and the Programa Severo Ochoa del Principado de Asturias; the Thalis and Aristeia programs cofinanced by EU-ESF and the Greek NSRF; the Rachadapisek Sompot Fund for Postdoctoral Fellowship, Chulalongkorn University and the Chulalongkorn Academic into Its 2nd Century Project Advancement Project (Thailand); the Kavli Foundation; the Nvidia Corporation; the SuperMicro Corporation; the Welch Foundation, contract C-1845; and the Weston Havens Foundation (USA).

\end{acknowledgments}

\bibliography{auto_generated}
\cleardoublepage \appendix\section{The CMS Collaboration \label{app:collab}}\begin{sloppypar}\hyphenpenalty=5000\widowpenalty=500\clubpenalty=5000\vskip\cmsinstskip
\textbf{Yerevan Physics Institute, Yerevan, Armenia}\\*[0pt]
A.M.~Sirunyan$^{\textrm{\dag}}$, A.~Tumasyan
\vskip\cmsinstskip
\textbf{Institut f\"{u}r Hochenergiephysik, Wien, Austria}\\*[0pt]
W.~Adam, F.~Ambrogi, T.~Bergauer, M.~Dragicevic, J.~Er\"{o}, A.~Escalante~Del~Valle, R.~Fr\"{u}hwirth\cmsAuthorMark{1}, M.~Jeitler\cmsAuthorMark{1}, N.~Krammer, L.~Lechner, D.~Liko, T.~Madlener, I.~Mikulec, F.M.~Pitters, N.~Rad, J.~Schieck\cmsAuthorMark{1}, R.~Sch\"{o}fbeck, M.~Spanring, S.~Templ, W.~Waltenberger, C.-E.~Wulz\cmsAuthorMark{1}, M.~Zarucki
% ...
\vskip\cmsinstskip
\dag: Deceased\\
1:  Also at Vienna University of Technology, Vienna, Austria\\
% ...
\end{sloppypar}
%%% END EDITABLE REGION %%%
\end{document}